\documentclass[traditabstract,a4]{aa} 
\usepackage{graphicx}
\usepackage{natbib}
\usepackage{txfonts}
\begin{document}
\title{A new extremely low-mass white dwarf in the NLTT catalogue\thanks{Based on 
observations collected at the European Organisation for Astronomical Research 
in the Southern Hemisphere, Chile under programme ID 82.D-0750.}}
\author{A. Kawka
        \and
        S. Vennes
}

\institute{Astronomick\'y \'ustav, Akademie v\v{e}d \v{C}esk\'e republiky, Fri\v{c}ova 298, CZ-251 65 Ond\v{r}ejov, Czech Republic\\
\email{kawka,vennes@sunstel.asu.cas.cz}
}

\date{Received; accepted}

\abstract{We report on the discovery of the extremely low-mass, hydrogen-rich 
white dwarf, NLTT 11748. Based on measurements of the effective temperature 
($8540\pm50$ K) and surface gravity ($\log{g} = 6.20\pm0.15$)
obtained by fitting the observed Balmer line profiles with synthetic spectra,
we derive a mass of $0.167\pm0.005\ M_\odot$. This object is one of only a 
handful of white dwarfs with masses below $0.2\ M_\odot$ that are believed to 
be the product of close binary evolution with an episode of Roche lobe overflow 
onto a degenerate companion (neutron star or white dwarf). Assuming 
membership in the halo population, as suggested by the kinematics and adopting 
a cooling age of $4.0 - 6.3$ Gyrs for the white dwarf, we infer a 
progenitor mass of $0.87 - 0.93\ M_\odot$. The likely companion has yet to be 
identified, but a search for radial velocity variations may help constrain its 
nature.}


\maketitle

\section{Introduction}

The average mass of white dwarfs is approximately $0.6\ M_\odot$, with the
bulk of these stars having a mass near this average value. Additional smaller
peaks at the ultramassive and low-mass ends of the mass distribution are 
observed. In the case of
low-mass white dwarfs ($M < 0.4\ M_\odot$), the galaxy is not old enough to
have produced such objects.
Several low-mass white dwarfs ($M < 0.45\ M_\odot$) have
been checked for radial velocity variations \citep[e.g.,][]{max2000,nel2005};
however, not all low-mass white dwarfs were found to be variable.
A statistical analysis of the incidence of binarity among low-mass white
dwarfs was conducted by \citet{nap2007} using the ESO SN Ia progenitor
survey, which included 26 white dwarfs with masses lower than 
$0.42\ M_\odot$. From this sample they find only 11 with variable radial
velocities showing that single low-mass white dwarfs do exist. 
It may be that some of these low-mass white dwarfs have indeed formed through
single-star evolution. \citet{kil2007c} argues that up to 50\% of low-mass
helium core white dwarfs $M \sim 0.4\ M_\odot$ could have formed from 
progenitors with supersolar metallicities.

Although some low-mass white dwarfs may evolve through single-star 
evolution, no single-star evolutionary path is known for extremely low-mass 
(ELM) white dwarfs ($M < 0.2\ M_\odot$), which
were initially identified
as companions to pulsars \citep[e.g.,][]{van1996,bas2006}.
Subsequently, several low-mass white dwarfs have been
discovered in the Sloan Digitial Sky Survey \citep{kle2004,eis2006},
with SDSS~J123410.37$-$022802.9 having the lowest mass at
$\sim 0.18\ M_\odot$ \citep{lie2004}. Other spectroscopic identifications 
include the
high-velocity white dwarf LP400-22, with a mass of $0.17\pm0.01\ M_\odot$
\citep{kaw2006}, and SDSS~J0917$+$46, which was discovered as part of a
search for hypervelocity B-type stars \citep{kil2007a}. Follow-up
observations of the last two of these stars using the 6.5m MMT telescope
found them to have variable
radial velocities. SDSS~J0917$+$46 was found to be in a binary with an
orbital period of $7.5936\pm0.0024$ hours \citep{kil2007b} and LP~400$-$22
was found to have an orbital period of approximately 1 day \citep{kil2009,ven2009}.
In both cases, the companion remains unknown, but a white
dwarf, neutron star, or a low-mass main-sequence star were considered as
possible companions.

The recent discovery of ELM white dwarfs offers new challenges for
binary evolution scenarios, as well as for calculating the internal structure of white dwarfs including the effect
of residual burning and model atmospheres
in a poorly explored range of stellar parameters.

We present our discovery and analysis of the ELM white dwarf NLTT~11748. 
Our spectroscopic and photometric observations are presented in \S 2. We
present our model atmosphere analysis of the object in \S 3 and discuss
our results in \S 4. We summarise in \S 5.

\section{Observations}

We obtained low-resolution spectroscopy of NLTT~11748 using the EFOSC2 (ESO 
Faint Object Spectrograph and Camera) attached to the New Technology Telescope 
(NTT) at La Silla Observatory. The 300 lines per mm grating with a blaze 
wavelength of 4000 \AA\ (Grism 11) was used to obtain two spectra with a
resolution of 16 \AA. The exposure time for each spectrum was 20 minutes.
The slitwidth was set to 1 arcsecond. We acquired both spectra
on UT 2008 October 21 in photometric conditions with the seeing
varying between 1 and 1.3 arcseconds throughout the night. A wavelength 
calibration spectrum of helium
and argon immediately followed the NLTT11748 spectra. The spectra 
were flux-calibrated using the flux standard Feige~110, which was obtained on 
the same night.

We then obtained ultraviolet photometry from the Galaxy Evolution Explorer 
({\it GALEX}) all sky survey. {\it GALEX} provides photometry in two bands, FUV 
and NUV. The FUV bandpass is $1344 - 1786$ \AA\ with an effective 
wavelength of 1528 \AA, and the NUV bandpass is $1771 - 2831$ \AA\ with an 
effective wavelength of 2271 \AA. We also obtained infrared photometry from 
the Two Micron All Sky Survey ({\it 2MASS}), which is available at VizieR at
the Centre de Donn\'ees astronomique (CDS)\footnote{http://cdsweb.u-strasbg.fr/CDS.html}. 
Table~\ref{tbl_phot} summarises the available photometry. The $V$ band 
photometry is estimated from our flux calibrated spectroscopy. We estimated
the uncertainty in the $V$ band to be 0.3 mag, which includes
systematic effects of varying seeing and data reduction procedures.
The {\it 2MASS} $K_S$ band is uncertain and is not considered further.

\begin{table}
\begin{minipage}{\columnwidth}
\caption{Photometry of NLTT~11748\label{tbl_phot}}
\centering
\renewcommand{\footnoterule}{}
\begin{tabular}{lcc}
\hline\hline
Band & Measurement & Model\footnote{Based on best model fit including $(E(B-V),R_V) = (0.10, 3.2)$. See \S 4.} \\
\hline
$FUV$ & $21.644\pm0.291$ & 21.839 \\
$NUV$ & $18.673\pm0.047$ & 18.494 \\
$V$   & $16.5\pm0.3$     & 16.493 \\
$J$   & $15.873\pm0.077$ & 15.933 \\
$H$   & $15.777\pm0.137$ & 15.768 \\
$K_S$ & $16.904$: & 15.700 \\
\hline
\end{tabular}
\end{minipage}
\end{table}

\section{Analysis}

\begin{figure}
\centering
\includegraphics[width=0.9\columnwidth]{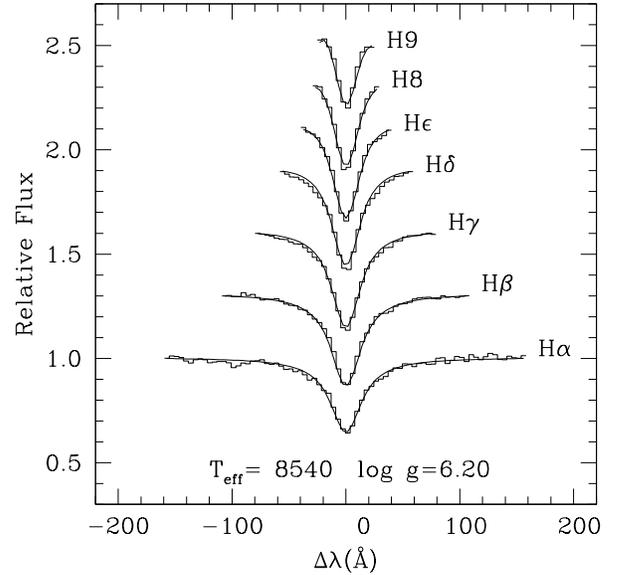}
\caption{Balmer line profiles of NLTT~11748 compared to the best-fit model 
spectrum with an effective temperature of $T_{\rm eff} = 8540\pm50$ K and 
a surface gravity of $\log{g} = 6.20\pm0.15$. \label{fig_fit}}
\end{figure}

We first fitted the observed Balmer lines with a grid of pure hydrogen synthetic
spectra that extends from 5000 to 7000 K (in steps of 500 K), 8000 to 
16\, 000 K (in steps of 1000 K), 18\, 000 to 32\, 000 K (in steps of 2000 K), and
36\, 000 to 100\, 000 K (in steps of 4000 K) at $\log{g} = 6.0$ to $9.5$ (in
steps of 0.25 dex). Following initial parameter estimations for NLTT~11748,
we calculated a denser grid with temperatures ranging from
8000 K to 9000 K (in steps of 100 K) and surface gravities ranging from
$\log{g} = 6.0$ to $7.0$ (in steps of 0.25 dex). Figure~\ref{fig_fit} shows 
the observed Balmer line 
profiles compared to the best-fit model spectrum. This model spectrum
corresponds to an effective temperature of $T_{\rm eff} = 8540\pm50$ K and
a surface gravity of $\log{g} = 6.20\pm0.15$. The quoted uncertainties are
only statistical ($1\sigma$). In general surface gravity measurements 
may be effected by a systematic offset between the 
assumed and the actual spectral resolution. 
At low surface gravity, a 1\AA\ offset would
correspond to $\approx 0.08$ dex shift in surface gravity or $0.007\ M_\odot$.

\begin{figure}
\centering
\includegraphics[width=0.9\columnwidth]{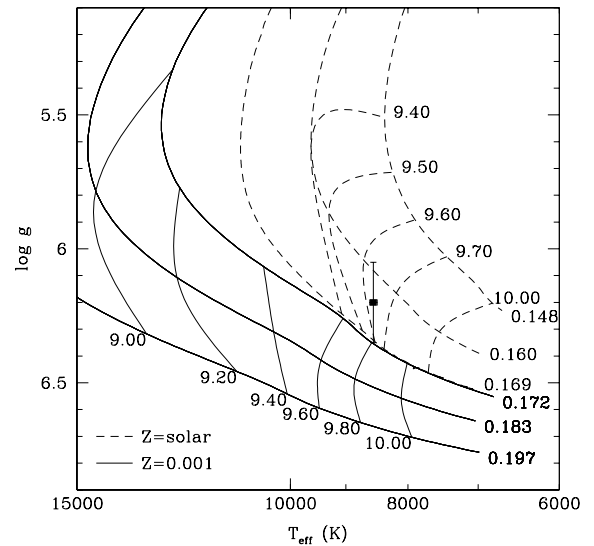}
\caption{The effective temperature and surface gravity of NLTT~11748 compared
to evolutionary mass-radius relations. The models with solar metallicity
and low metallicity ($Z=0.001$) are from \citet{ser2001} and \citet{ser2002}, 
respectively. The isochrones are labelled with $\log{({\rm years})}$ and the
curves of constant mass are in units of $M_\odot$.
\label{fig_evol}}
\end{figure}

Using the evolutionary mass-radius relations of
\citet{ser2002}, we obtain a mass of $0.167\pm0.005\ M_\odot$ and a cooling
age of $4.0 - 6.3$ Gyrs. The use of low-metallicity models from \citet{ser2002} 
is justified by the old age of the star. We also calculated the absolute
magnitude ($M_V$) using these models.
For comparison we also calculated a mass of $0.164\pm0.005\ M_\odot$ 
using the solar metallicity models of \citet{ser2001}, which is very
similar to the low-metallicity mass determination. However, the cooling age of
the system assuming solar metallicity would be between 1 and 4 Gyrs, which is 
significantly lower than the cooling age predicted by the low-metallicity 
models. The difference in the cooling age can be attributed to the effect of
metallicity on the residual hydrogen burning, which is expected for white
dwarfs with such a low mass with solar or low-metallicity progenitors \citep{ser2002}.
Figure~\ref{fig_evol} shows the effective temperature and surface gravity of 
NLTT~11748
compared to the evolutionary mass-radius relations of
\citet{ser2001} and \citet{ser2002}. 
Table~\ref{tbl_prop} lists our adopted values for $T_{\rm eff}$, $\log{g}$,
$M$, and $M_V$. 

\section{Discussion}

\subsection{Kinematics}

We selected NLTT~11748 (LP413$-$40) as a white dwarf candidate from the 
revised NLTT catalogue \citep{sal2003} using the reduced proper motion diagram 
\citep{sal2002,kaw2004}. \citet{luy1977} originally listed NLTT~11748 in his 
white dwarf catalogue, and, more recently the star has also been recovered in the 
LSPM catalogue as LSPM~J0345$+$1748 \citep{lep2005}.
The rNLTT and LSPM catalogue remeasured the proper motions independently. We 
list the weighted average of these measurements in Table~\ref{tbl_prop}. We also
measured the radial velocity of the star using the Balmer line series 
(H$\alpha$ to H$\gamma$) and the corresponding heliocentric velocity is 
provided in Table~\ref{tbl_prop}.

The photometric distance estimate toward NLTT~11748 is possibly effected by 
extinction in the line of sight.
NLTT~11748 is near the Taurus-Auriga molecular cloud, which is a well known 
star formation region located at a distance of $\sim 140$ pc \citep[e.g.,][]{li1998}.
The extinction in the line of sight toward NLTT~11748 based on the maps of
\citet{sch1998} is $E(B-V) = 0.2828$, which
corresponds to a large extinction in the ultraviolet. Assuming $R_V = 3.2$
and using the relations from \citet{car1989}, we estimate the extinction
in the  NUV and FUV bands to be $A_{\rm NUV} = 2.6$ and $A_{\rm FUV} = 2.3$,
respectively. However, we do not expect such a large extinction because the 
object is still well within the plane. 

\begin{figure}
\centering
\includegraphics[width=\columnwidth]{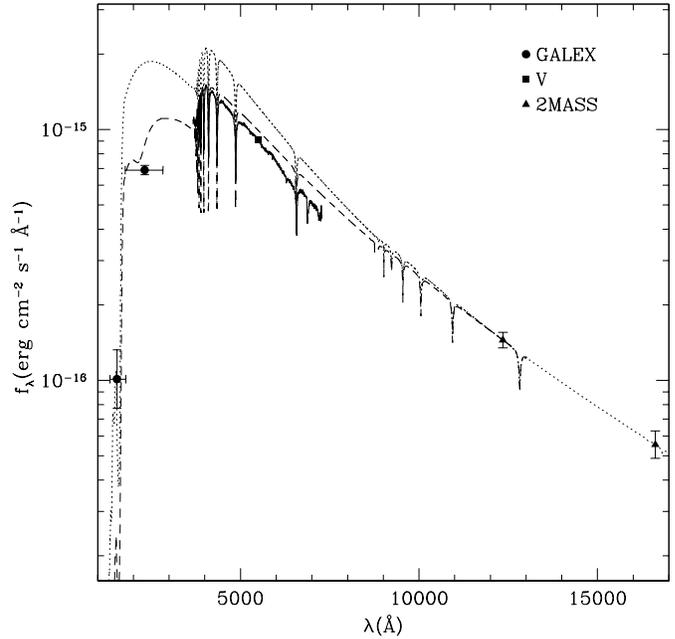}
\caption{Energy distribution of NLTT~11748 combining all available data 
compared to our hydrogen-rich spectrum at $T_{\rm eff} = 8500$ K and
$\log{g} = 6.25$ without extinction (dotted line) and with extinction assuming
$E(B-V) = 0.10$ (dashed line). The filled circles are the observed 
data, and the empty circles the predicted photometric points if we assume 
no extinction.
\label{fig_sed}}
\end{figure}

The extinction in the infrared should be very small, so we calculated
a distance of $199\pm40$ pc using the $J$ magnitude. Using the effective
temperature and surface gravity from the spectral fit, we calculated the
absolute magnitude in the $FUV$, $NUV$, $V$, $J$, and $H$ bands. The expected $NUV - J$ 
for the white dwarf is $1.90$, but the observed $NUV - J$ colour is 
$2.836\pm0.090$. This corresponds to an extinction of $A_{NUV} = 0.935$ in
the NUV and $E(B-V) \approx 0.10$. Using the synthetic spectrum of the
white dwarf at the derived temperature and surface gravity, we calculated a 
synthetic spectrum taking extinction into account where $E(B-V) = 0.10$ and
$R_V = 3.2$.
Figure~\ref{fig_sed} shows the observed energy distribution compared to a 
synthetic spectrum with and without extinction. The corresponding model 
magnitudes including extinction are given in Table~\ref{tbl_phot}.

\begin{figure}[t]
\centering
\includegraphics[width=\columnwidth]{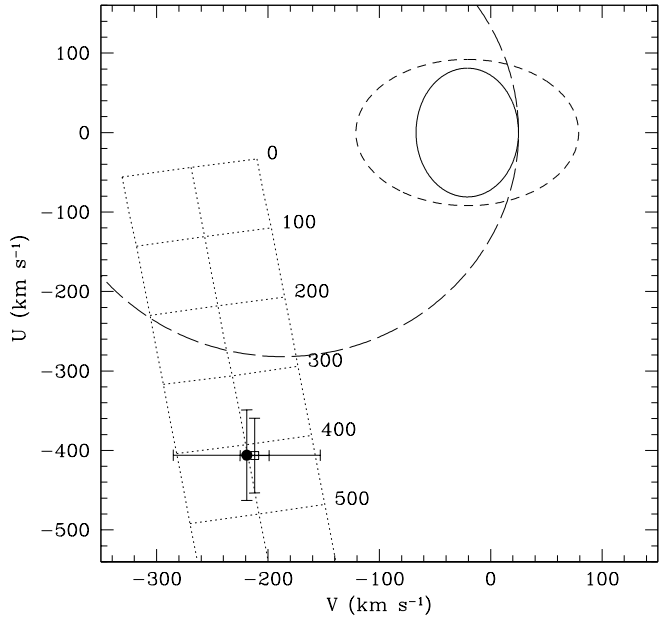}
\caption{The $U$ and $V$ velocities of NLTT~11748 ({\it full circle}) and 
LP~400-22 ({\it open square}) compared to the $2\sigma$
thin- ({\it full line}) and thick- ({\it short dashed line}) disk ellipses and
the $2\sigma$ halo ellipse ({\it long dashed line}). The predicted $UV$ 
velocities ({\it dotted line}) for varying radial velocity measurements are 
shown with the assumed velocity.
\label{fig_UV}}
\end{figure}

Based on the proper motion, distance, and velocity measurements, we calculated
the kinematics (Table~\ref{tbl_prop}) following \citet{joh1987} and assuming
that the solar motion relative to the local standard of rest (LSR) is 
$(U_\odot, V_\odot, W_\odot) = (10.1, 4.0, 6.7)$ as determined by \citet{hog2005}. 
Figure~\ref{fig_UV} shows the $U$ and $V$ velocities of NLTT~11748 assuming
a radial velocity of $415$ km~s$^{-1}$. However, NLTT~11748 probably resides
in a close binary, and the correct systemic velocity still needs to be
measured\footnote{The measured velocity difference between the two 
available spectra is $\sim 40$ km~s$^{-1}$, which suggests the star is showing 
velocity variations, but the $1\sigma$ uncertainties of the two 
measurements overlap.}.
We illustrate the possible variations in $UV$ velocities assuming a range of
values from 0 to 600 km~s$^{-1}$ for the systemic velocity.
NLTT~11748 bears a striking resemblance with the other high-velocity ELM white
dwarf LP~400-22 \citep{kil2009,ven2009}.

\begin{table}
\caption{Properties of NLTT~11748\label{tbl_prop}}
\centering
\begin{tabular}{llc}
\hline\hline
Parameters & Measurement & Reference \\
\hline
R.A. (J2000.0)        & 03 45 16.83 & 1 \\
Dec. (J2000.0)        & +17 48 08.71 & 1 \\
Effective temperature & $8540\pm50$ K & 2 \\
Surface gravity       & $6.20\pm0.15$ & 2 \\
Mass                  & $0.167\pm0.005\ M_\odot$ & 2 \\
$M_V$                 & $9.72\pm0.33$ & 2 \\
Distance              & $199\pm40$ pc & 2 \\
$\varv_{\rm rad}$     & $415\pm50$ km s$^{-1}$ & 2 \\
Proper motion         & $\mu_{\rm RA} = 0.2361\pm0.0043\arcsec$ yr$^{-1}$ & 1,3 \\
                      & $\mu_{\rm Dec} = -0.1792\pm0.0043\arcsec$ yr$^{-1}$ & 1,3 \\
Kinematics            & $U = -406\pm57$ km s$^{-1}$ & 2 \\
                      & $V = -219\pm66$ km s$^{-1}$ & 2 \\
		      & $W = -161\pm35$ km s$^{-1}$ & 2 \\
\hline               
\end{tabular}
References: (1) \citet{sal2003}; (2) This work; (3) \citet{lep2005}
\end{table}

\subsection{Binary evolution}

Given that the two field ELM white dwarfs (LP~400-22, SDSS J0917$+$46) that were checked for
radial velocity variations were found to be in a binary system, it is 
highly probable that NLTT~11748 is also in a binary system. In 
the two confirmed field systems, the companion remains unknown, although the
other known ELM white dwarfs are companions to pulsars \citep{van1996,bas2006},
so it is possible that NLTT~11748 has a neutron star companion.
If we assume that the companion is a neutron star, then the orbital period
is expected to be $\sim 10 - 15$ hours based on the
mass versus orbital period relations for He WDs with a neutron star companion
\citep{ben2005}. We searched for radio sources in the vicinity of 
NLTT~11748 using VizieR, and the nearest source is an unrelated object 2.5 
arcminutes away.

It is also possible that the companion is another, more massive white
dwarf. \citet{nel2001} present a population synthesis of double
degenerate systems that experience at least two phases of mass transfer,
out of which at least one will result in a common envelope allowing the
white dwarfs to spiral in. Their population synthesis also shows a
correlation between the mass of the brighter white dwarf and the orbital period.

Although unlikely, the companion could be a low-mass main sequence star. Our 
optical spectrum and 2MASS photometry do not show any evidence of a cool 
companion. Since NLTT~11748 is relatively faint,
2MASS provides a reliable measurement in the $J$ band, with $H$ having a 
considerably greater uncertainty. The absolute $J$ magnitude for the white 
dwarf is $M_J = 9.3$. A red dwarf of spectral type M5 to M5.5 would 
have equally contributed
in this spectral region \citep{kir1994}, so a prospective 
late type companion would have to be much less massive than $0.2\ M_\odot$.

Evolutionary scenarios that produce ELM white dwarfs involve an episode of
Roche lobe overflow and accretion onto a degenerate companion that stripped
the envelope of the white dwarf progenitor leaving a low-mass degenerate core
\citep{wil2002,ben2005}. Assuming halo membership, as suggested by the 
kinematics, the progenitor lifetime on the main-sequence would range from 
7 to 9 Gyrs corresponding to a late G star with a mass of 
$0.87 - 0.93\ M_\odot$ \citep{gir2000}.

\section{Conclusions}

Based on a model atmosphere analysis, we have shown that NLTT~11748 is an ELM 
white dwarf ($M = 0.167\pm0.005\ M_\odot$). Although we assumed a 
pure hydrogen composition, near ultraviolet spectroscopy is required to confirm 
our model atmosphere analysis and to constrain the abundance of heavy elements 
in the atmosphere. NLTT~11748 is one of only a handful of low-gravity white 
dwarfs that allow a test of our models in these unique conditions.

The star almost certainly belongs to a close binary system with a still
unidentified degenerate companion. A probable evolutionary scenario involves a 
post supernova system comprising an evolving G star and a neutron star that
experienced Roche lobe overflow accretion, leaving an ELM degenerate star. 
This object would be part of a growing family of low-mass X-ray binary remnants. 
A radial velocity study should be able to constrain the orbital period and the 
mass function of the unseen companion. On the other hand, the binary companion 
may be another white dwarf, thereby confirming another formation 
channel for ELM white dwarfs in parallel with the channel involving a neutron 
star. Our kinematic study shows that the system belongs to the Galactic halo,
which helps constrain the age, hence the mass of the donor star.

Finally, studies of ELM white dwarfs are useful in exploring the effect of 
residual burning on white dwarf mass-radius relations. In this regard an 
independent estimate of the radius using parallax measurements is essential.
In addition, as more ELM white dwarfs are discovered, a more comprehensive 
set of evolutionary tracks for these stars is required.

\begin{acknowledgements}

A.K. and S.V. are supported by grants IAA301630901 and IAA300030908 from the GA
AV, respectively. A.K. also acknowledges support from the Centre for Theoretical
Astrophysics (LC06014). This research has made use of the VizieR catalogue 
access tool, CDS, Strasbourg, France.
This publication makes use of data products from the 2MASS,
which is a joint project of the University of Massachusetts and the Infrared 
Processing and Analysis Center/California Institute of Technology, funded by 
the National Aeronautics and Space Administration and the National Science 
Foundation.

\end{acknowledgements}

\end{document}